\documentclass[final,5p,times,twocolumn]{elsarticle}

\usepackage{amsmath}
\usepackage{amssymb}
\usepackage{algorithm}
\usepackage{algorithmic}
\usepackage{bm} 
\usepackage{booktabs}
\usepackage{color}
\usepackage{changepage}
\usepackage{epsfig}
\usepackage[T1]{fontenc}
\usepackage{graphicx}
\usepackage{lipsum}
\usepackage{lineno,hyperref}
\usepackage{multirow}
\usepackage{pbox}
\usepackage{subfig}
\usepackage{tabularx}

\usepackage{tabularx}
\usepackage{url}  
\usepackage{verbatim}          

\newlength{\offsetpage}
	{\end{adjustwidth}}
\modulolinenumbers[5]

\journal{Elsevier}

\begin{document}
	
\begin{frontmatter}
	
	\title{PyMIC: A deep learning toolkit for annotation-efficient medical image segmentation}
	
	
	\author[address_uestc,address_pjlab]{Guotai Wang\corref{mycorrespondingauthor}}
	\cortext[mycorrespondingauthor]{Corresponding author}
	\ead{guotai.wang@uestc.edu.cn}
	
	\author[address_uestc,address_pjlab]{Xiangde Luo}
	\author[address_uestc]{Ran Gu}
	\author[address_jhu]{Shuojue~Yang}
	\author[address_uestc]{Yijie Qu}
	\author[address_uestc]{Shuwei~Zhai}
	\author[address_uestc]{Qianfei~Zhao}
	\author[address_scu]{Kang~Li}
	\author[address_uestc,address_pjlab]{Shaoting~Zhang}
	
	
	
	\address[address_uestc]{School of Mechanical and Electrical Engineering, University of Electronic Science and Technology of China, Chengdu, China}
	\address[address_pjlab]{Shanghai Artificial Intelligence Laboratory, Shanghai, China}
	\address[address_jhu]{Laboratory for Computational Sensing and Robotics, Johns Hopkins University, Baltimore, USA}
	\address[address_scu]{West China Biomedical Big Data Center, West China Hospital, Sichuan University, Chengdu, China}
	
	\begin{abstract}
		Background and Objective: Open-source deep learning toolkits are one of the driving forces for developing medical image segmentation models. Existing toolkits mainly focus on fully supervised segmentation and require full and accurate pixel-level annotations that are time-consuming and difficult to acquire for segmentation tasks, which makes learning from imperfect labels highly desired for reducing the annotation cost. We aim to develop a new deep learning toolkit to support annotation-efficient learning for medical image segmentation. 
		
		Methods: Our proposed toolkit named PyMIC is a modular deep learning library for medical image segmentation tasks. In addition to basic components that support development of high-performance models for fully supervised segmentation, it contains several advanced components tailored for learning from imperfect annotations, such as loading annotated and unannounced images, loss functions for unannotated, partially or inaccurately annotated images, and training procedures for co-learning between multiple networks, etc. PyMIC supports development of semi-supervised, weakly supervised and noise-robust learning methods for medical image segmentation.
		
		Results: We present several illustrative medical image segmentation tasks based on PyMIC: (1) Achieving competitive performance on fully supervised learning; (2) Semi-supervised cardiac structure segmentation with only 10\% training images annotated; (3) Weakly supervised segmentation using scribble annotations; and (4) Learning from noisy labels for chest radiograph segmentation.
		
		Conclusions: The PyMIC toolkit is easy to use and facilitates efficient development of medical image segmentation models with imperfect annotations. It is modular and flexible, which enables researchers
		to develop high-performance models with low annotation cost. The source code is available at:\url{https://github.com/HiLab-git/PyMIC}.
		
	\end{abstract}
	
	
\end{frontmatter}


\section{Introduction}
Automatic segmentation of medical images is essential for computer-assisted  diagnosis and treatment planning/guidance of disease~\cite{Shen2017}. Recently, with the fast development of deep learning techniques, the segmentation performance has been largely improved and even becomes comparable to that of human experts for clinical use~\cite{Tang2019}. In addition to hardware revolution and increased data availability, open-source deep learning toolkits are one of the driving forces for
the boom of deep learning methods and their applications in the medical image segmentation  community~\cite{Isensee2021,monai2022}. 
Existing general-purpose deep learning frameworks such as TensorFlow~\cite{Abadi2016}, PyTorch~\cite{Paszke2019} and Keras\footnote{https://keras.io/} provide modular and lower-level functions for deep learning in a wider community of computer vision and natural language processing, without specific functionalities for medical image segmentation, and adapting them for medical images requires substantial implementation efforts.

There have been several open-source projects proposed for medical image computing including segmentation tasks in the last couple of decades. Some early toolkits implemented traditional algorithms for medical image computing, such as NiftyReg~\cite{Modat2010a} and Elastix~\cite{Klein2010a} for registration and NiftySeg~\cite{Cardoso2012} for segmentation. Researchers also developed some application-specific toolkits, such as Freesurfer~\cite{fischl2012} and FSL~\cite{Smith2004} for structural and functional neuroimaging analysis. However, they did not support deep learning models for medical image analysis. In recent years, several research groups have extended general deep learning toolkits with more domain-specific functionalities to ease medical image computing based on deep learning. NiftyNet~\cite{Gibson2018} and DLTK~\cite{Pawlowski2017} enable fast development of deep learning models for medical image analysis using TensorFlow~\cite{Abadi2016}. However, they are not actively maintained and updated now. DeepNeuro~\cite{Beers2021} is also a framework built on the top of TensorFlow, and provides a templating language for designing and implementing deep learning pipelines for neuroimaging. 
MONAI (Medical Open Network for AI)~\cite{monai2022} 
that is the predecessor of  NiftyNet and built on PyTorch provides modules for data loading, training paradigm, network structure and loss functions for medical image computing models. TorchIO~\cite{Perez-Garcia2021} is a library for data loading, prepossessing, augmentation and patch sampling of medical images in deep learning. Pymia~\cite{Jungo2021} also provides functionalities for handling  medical images in different formats, in addition to evaluation of deep learning models. It can be integrated into both TensorFlow and PyTorch pipelines. The nnU-Net~\cite{Isensee2021} framework based on PyTorch is specifically designed for deep learning-based medical image segmentation, and it automatically configures itself for data preprocessing, network architecture, training and post-processing, leading to state-of-the-art performance on many international biomedical image segmentation challenges. 

However, the existing toolkits only considered fully supervised learning for medical image segmentation tasks, with the assumption that pixel-level dense and accurate annotations are available for the training set. The implementation of such a supervised learning paradigm in these toolkits is simple and straightforward: loading image and label pairs from a fully annotated dataset and training a single network with common fully-supervised loss functions. For medical image segmentation, the acquisition of pixel-level  annotations is extremely time-consuming, labor-insensitive and  difficult with high cost~\cite{Tajbakhsh2019}. First, differently from natural images, annotation of medical images requires many expertise, where the availability of experts will limit the annotation process. Second, medical images such as Computed Tomography (CT) and Magnetic Resonance Imaging (MRI) are volumetric data often with hundreds of 2D slices, and delineating multiple organs~\cite{Tang2019} or lesions in a volume would take several hours, making annotating a large dataset very expensive and time-consuming. Thirdly, due to the low contrast of medical images and varying experience of annotators, the annotation may be inaccurate and contain some noise~\cite{Wang2020c}.  The standard supervised learning paradigm in existing toolkits do not provide support for dealing with such datasets with incomplete or imperfect annotations, which impedes researchers from fast developing their models when the annotations are limited.

In recent years, annotation-efficient learning for medical image segmentation has attracted increasing attention as it can largely reduce the annotation cost for training deep learning models~\cite{Tajbakhsh2019}. For example, Semi-Supervised Learning (SSL) leverages a small set of annotated images and a large set of unannotated images for training, which avoids annotating the entire dataset. Some Weakly Supervised Learning (WSL) methods only require sparse or coarse annotations (e.g., scribble 
and bounding box 
that are much cheaper than pixel-level annotations. In addition, some researchers use noisy annotations given by non-experts for training when access to experts is limited~\cite{Wang2020c,Zhang2020trinet}, i.e., Noisy Label Learning (NLL). 

The implementation of such methods is largely different from that of supervised learning, and requires a lot of advanced functionalities for data handling, network design, loss function and training pipeline that are not accessible in existing toolkits. For example, semi-supervised learning methods often load annotated images and unannotated images in a single batch for training~\cite{Luo2022a,Yu2019}, which requires redesign of the data loader and batch sampler. The loss function in semi- and weakly supervised learning is often a combination of supervised loss for annotated pixels and different types of regularization for unannotated pixels~\cite{Liu2022pr,Luo2022a}, and such regularization and their combination are ignored in existing toolkits. In addition, modification of network structures with multiple branches or co-learning between different networks~\cite{Zhang2020trinet,Yu2019,Wang2021d} are often involved in annotation-efficient learning. Therefore, a lot of implementation efforts are needed by these algorithms for learning from imperfect annotations. As independent reimplementation of all of these customized modules leads to substantial duplication of efforts and inhibits fair comparison between competing methods, a unified software toolkit with built-in support for annotation-efficient learning is highly desired.

This work presents the open-source PyMIC (PyTorch for Medical Image Computing) 
toolkit based on PyTorch for annotation-efficient learning for medical image segmentation. In addition to flexible and modular functionalities for training, inference and evaluation of medical image segmentation models, it provides advanced built-in modules for data handling, network design,  loss functions and training procedures required by semi-supervised, weakly supervised and noisy label learning. It allows fast development of deep learning segmentation models using limited annotations, and facilitates comparison between different methods.

\section{Methods}
The PyMIC toolkit aims to fill the gap between current deep learning toolkits for fully supervised learning and learning with limited (e.g., partial, sparse and noisy) annotations in medical image segmentation. PyMIC is built on PyTorch that provides basic ingredients for deep learning but without specific functionalities for dealing with medical images and annotation-efficient learning. PyMIC implements higher-level interfaces for medical image segmentation with advanced functionalities optimized for learning from imperfect annotations.

\subsection{System overview}
\begin{figure*}[t]
	\centering
	\includegraphics[width=1.0\linewidth]{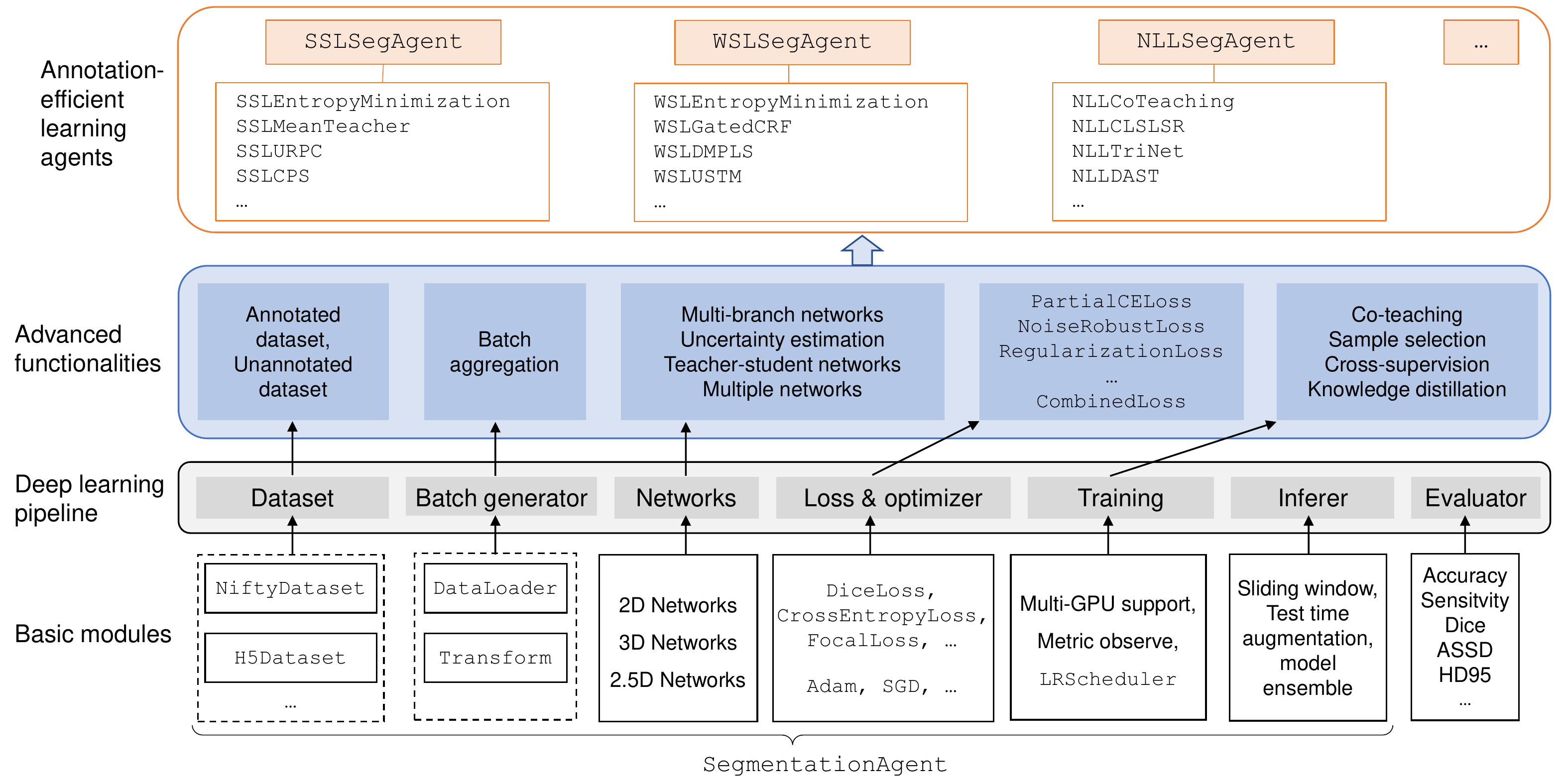}
	\caption{ An overview of the PyMIC framework for medical image segmentation with annotation-efficient learning.}
	\label{fig:overview}
\end{figure*}
Fig.~\ref{fig:overview} shows an overview of PyMIC that contains several modular components. Firstly, it leverages lower-level functions in PyTroch to implement a  modularized and simplified deep learning pipeline, and defines a 
{\fontfamily{cmtt}\selectfont SegmentationAgent} to connect and encapsulate  several basic core components for training and inference of deep learning models: {\fontfamily{cmtt}\selectfont
Dataset, BatchGenerator} and {\fontfamily{cmtt}\selectfont Network} for common dataflow, {\fontfamily{cmtt}\selectfont Loss} and {\fontfamily{cmtt}\selectfont
Optimizer} used in {\fontfamily{cmtt}\selectfont Training} procedure and an  {\fontfamily{cmtt}\selectfont Inferer} for inference. The  {\fontfamily{cmtt}\selectfont SegmentationAgent} can easily handle supervised learning tasks, and accepts configuration files to run the deep learning pipeline with different settings of datasets, data augmentation strategies, networks and loss functions, etc. 

Secondly, we extend the basic core components with advanced functionalities that are specifically designed for handling and training with datasets with imperfect annotations, such as batch aggregation that aggregates samples from annotated and unannotated images into a single batch, multi-branch networks, uncertainty estimation and regularization losses used in semi- and weakly supervised segmentation. We also provide advanced training procedures such as co-learning between different networks simultaneously and sample selection for dynamically filtering out samples with low-quality annotations.   

Thirdly, based on {\fontfamily{cmtt}\selectfont SegmentationAgent} and the advanced functionalities, we define derived classes for annotation-efficient learning agents including {\fontfamily{cmtt}\selectfont SSLSegAgent},  {\fontfamily{cmtt}\selectfont WSLSegAgent} and {\fontfamily{cmtt}\selectfont NLLSegAgent} that provide general frameworks for semi-supervised learning, weakly supervised learning and noisy label learning for segmentation, respectively. As shown in Fig.~\ref{fig:overview}, these classes are parent classes of specific annotation-efficient learning algorithms, such as {\fontfamily{cmtt}\selectfont SSLUAMT} that uses Uncertainty-Aware Mean Teacher (UAMT) for semi-supervised learning~\cite{Yu2019} and {\fontfamily{cmtt}\selectfont WSLGatedCRF} that uses gated Conditional Random Field (CRF) for weakly supervised learning~\cite{Obukhov2019}. In the following subsections, we give detailed description of these modules respectively.

\subsection{Basic modules for fully supervised learning}\label{sec:method_full}
PyMIC implements several core components of a deep learning pipeline that are specialized for medical image segmentation and encapsulated into the   {\fontfamily{cmtt}\selectfont SegmentationAgent} class. These components include dataset and batch generator, network, loss and optimizer, training and inference procedures, etc. An {\fontfamily{cmtt}\selectfont Evaluator} is also implemented for quantitative evaluation of segmentation results with various evaluation metrics.

Medical images such as CT and MRI are saved in special formats such as DICOM, Nifti and Analyze that are not supported by general deep learning frameworks including Tensorflow and PyTorch. The Nifti format is the most widely used in the community, and the other formats can be converted to Nifti that saves image data array together with metadata including spatial information, patient information and acquisition information. Therefore, we define a {\fontfamily{cmtt}\selectfont NiftyDataset} to load medical images, which uses {\fontfamily{cmtt}\selectfont load\_image\_as\_nd\_array()} as a general interface to read different types of images as $N$-dimensional arrays with meta information including pixel spacing. For example, a 3D image is loaded as a 4D array with the shape of $C\times D \times H \times W$, where  $C$, $D$, $H$ and $W$ represent the channel number, slice number, height and width, respectively. Note that this function can load a regular 2D image with different formats such as .jpg, .png and .bmp as a 3D array with the shape of $C\times H\times W$ as well. In addition, to load large volumes efficiently, PyMIC defines a {\fontfamily{cmtt}\selectfont H5Dataset} to support HDF5 (hierarchical data format version 5) that allows access of chunks of data without loading the entire volume. 

With {\fontfamily{cmtt}\selectfont NiftyDataset} or {\fontfamily{cmtt}\selectfont H5Dataset},  {\fontfamily{cmtt}\selectfont BatchGenerator} is implemented to sample a batch from the dataset. It reuses {\fontfamily{cmtt}\selectfont DataLoader} in PyTorch and  {\fontfamily{cmtt}\selectfont Transform} for data preprocessing and augmentation. {\fontfamily{cmtt}\selectfont Transform} includes a set of intensity and spatial transforms, such as {\fontfamily{cmtt}\selectfont NormalizeWithMeanStd} and {\fontfamily{cmtt}\selectfont NormalizeWithMinMax} for intensity normalization, {\fontfamily{cmtt}\selectfont GammaCorrection} and {\fontfamily{cmtt}\selectfont GaussianNoise} for random intensity augmentation and {\fontfamily{cmtt}\selectfont RandomCrop}, {\fontfamily{cmtt}\selectfont RandomFlip}, {\fontfamily{cmtt}\selectfont RandomRescale},
{\fontfamily{cmtt}\selectfont RandomRotate}, {\fontfamily{cmtt}\selectfont Pad} and others for spatial augmentation. Note that these transforms support 2D and 3D images in PyMIC, and the inverse operation is allowed for spatial transforms.

Several types of neural networks are defined in PyMIC for medical image segmentation. 2D networks such as {\fontfamily{cmtt}\selectfont UNet2D}~\cite{Ronneberger2015}, {\fontfamily{cmtt}\selectfont AttentionUNet2D}~\cite{Oktay2018} and {\fontfamily{cmtt}\selectfont NestedUNet2D}~\cite{Zhou2018} are implemented to segment 2D images or slices of 3D images. Some 3D networks such as {\fontfamily{cmtt}\selectfont UNet3D}~\cite{Abdulkadir2016} and  {\fontfamily{cmtt}\selectfont UNet3D\_ScSE}~\cite{Roy2019} are available for segmentation of 3D volumes. In addition, many medical image datasets have high intra-plane resolution and low inter-plane resolution, making networks using purely 2D or 3D convolutions less effective. PyMIC supports 2.5D networks such as {\fontfamily{cmtt}\selectfont UNet2D5} that combines 2D and 3D convolutions in a single network to better deal with such images. What's more, PyMIC supports several loss functions widely used for image segmentation, such as {\fontfamily{cmtt}\selectfont DiceLoss}, {\fontfamily{cmtt}\selectfont CrossEntropyLoss}, {\fontfamily{cmtt}\selectfont FocalLoss} and {\fontfamily{cmtt}\selectfont ExpLogLoss}. 

During training, PyMIC reuses built-in optimizers and Learning Rate (LR) scheculers in PyTorch. 
It also allows training with multiple GPUs by reusing {\fontfamily{cmtt}\selectfont DataParallel} that parallelizes the network by splitting the input across a list of devices. In addition, PyMIC will measure the segmentation performance via some evaluation metrics such as Dice on the training and validation sets during training, which is better than just using the loss values for selecting the best checkpoint. 

For inference, an {\fontfamily{cmtt}\selectfont Inferer} object is defined in PyMIC with functionalities for medical images that are not readily available in PyTorch. It allows sliding window inference for a large volume by automatically spiting the input into a list of patches and aggregating the corresponding results into a single segmentation volume in the original image space. Test time augmentation is also supported by {\fontfamily{cmtt}\selectfont Inferer}, where different transformed versions of the input are segmented and transformed inversely to the original spatial coordinates for segmentation fusion. In addition, {\fontfamily{cmtt}\selectfont Inferer} allows ensemble of multiple checkpoints for higher segmentation accuracy and robustness.  

With the segmentation ground truth and network's prediction, an {\fontfamily{cmtt}\selectfont Evaluator} object allows quantitative evaluation of the segmentation results. It supports a range of metrics including sensitivity, specificity, accuracy, precision, Dice coefficient, Average Symmetric Surface distance (ASSD), Hausdorff Distance (HD) and 95 percentile of HD (HD95), etc. Users can specify the list of classes and the names of metrics for evaluation, and the results are saved in .csv format via a  {\fontfamily{cmtt}\selectfont CSVWriter}. {\fontfamily{cmtt}\selectfont Evaluator} also allows metric observation during training.

\subsection{Advanced functionalities for learning from imperfect annotations}
Based on the above modules that facilitate fully supervised learning for segmentation, PyMIC further extends them in several aspects  with advanced functionalities to support annotation-efficient learning, including data handling, networks, loss functions and training procedures specified for semi-supervised, weakly supervised and noisy-label learning. 

For semi-supervised setting where the dataset consists of an annotated subset and an unannotated subset, {\fontfamily{cmtt}\selectfont NiftyDataset} and {\fontfamily{cmtt}\selectfont H5Dataset} are extended by allowing loading images without labels. In addition, two instances of such datasets are created in the segmentation agent, and they are for the annotated and unannotated subsets, respectively. SSL methods usually require annotated and unannotated images in a single batch~\cite{Luo2022a,Yu2019}, and PyMIC supports this by loading $N_L$ and $N_U$ images from the two dataset instances respectively to compose a batch with $N_L + N_U$ images, i.e., batch aggregation.

PyMIC supports several advanced and specialized networks for annotation-efficient learning. For example, multi-branch networks are often used for SSL~\cite{Luo2022a,Ouali2020}, WSL~\cite{Luo2022miccai} and NLL~\cite{Yang2022jbhi} tasks, and PyMIC extends encoder-decoder networks like U-Net with multiple decoders or multi-level predictions for this purpose. Different uncertainty estimation methods are also supported in PyMIC, e.g., Monte Carlo Dropout~\cite{Yu2019}, test-time augmentation and multi-branch predictions~\cite{Luo2022a} that are commonly used for SSL tasks. In addition, PyMIC has implementations of self-ensembling of networks that are widely used in Mean Teacher-based methods~\cite{Yu2019,Liu2022pr}, and it extends the {\fontfamily{cmtt}\selectfont SegmentationAgent} to allow multiple instances of networks that are required in some SSL~\cite{Chen2021cps} and NLL~\cite{Zhang2020trinet,Han2018} methods.

In addition to general supervised loss functions, PyMIC has implementations of several losses specialized for annotation-efficient learning. For example, in SSL and WSL, the predictions for unannotated pixels are often regularized. The {\fontfamily{cmtt}\selectfont RegularizationLoss} is implemented to support different kinds of regularization, such as entropy-minimization, total variation, and size and shape regularization, etc.  {\fontfamily{cmtt}\selectfont NoiseRobustLoss} is implemented for dealing with noisy labels, such as noise-robust Dice loss~\cite{Wang2020c}, Generalized Cross Entropy (GCE) loss~\cite{Zhang2018d} and Mean Absolute Value (MAE) loss~\cite{Ghosh2017}, etc. In addition, to facilitate the combination of different kinds of loss functions, a {\fontfamily{cmtt}\selectfont CombinedLoss} allows users to select the loss functions for combination and specify their corresponding weights.

The {\fontfamily{cmtt}\selectfont Training} procedure of {\fontfamily{cmtt}\selectfont SegmentationAgent} is derived to support different training paradigms for annotation-efficient learning. For example, in co-teaching~\cite{Han2018}, the training involves two networks where each network selects samples based on loss values for the other network in a batch. In Mean Teacher~\cite{Tarvainen2017}, a self-ensembling of the student network serves as a teacher the output of which is used to supervise the student. {\fontfamily{cmtt}\selectfont NLLCoTeaching} and {\fontfamily{cmtt}\selectfont SLLMeanTeacher} are derived from {\fontfamily{cmtt}\selectfont SegmentationAgent} by redefining the training process to support these two training paradigms, respectively. Other types of training paradigms, such as knowledge distillation and  confidence learning are also supported in similar ways in PyMIC.

\subsection{Annotation-efficient leaning agents}
\begin{table*}
	\centering
	\small 
	\caption{Overview of annotation-efficient learning methods currently implemented in PyMIC for medical image segmentation. SSL, WSL and NLL represent semi-supervised learning, weakly supervised learning and noisy label learning, respectively. }
	\begin{tabular}{l|l|l|l}
		\hline
		Category & Method & Reference & Remarks \\  \hline
		SSL  &  {\fontfamily{cmtt}\selectfont SSLEntropyMinimization} & Grandvalet et al., 2005~\cite{Grandvalet2005} & Originally proposed for classification \\
		&  {\fontfamily{cmtt}\selectfont SSLMeanTeacher} & Tarvainen et al., 2017~\cite{Tarvainen2017} & Originally proposed for classification \\
		&  {\fontfamily{cmtt}\selectfont SSLUAMT} & Yu et al., 2019~\cite{Yu2019} & Uncertainty-aware mean teacher \\
		&  {\fontfamily{cmtt}\selectfont SSLURPC} & Luo et al., 2022~\cite{Luo2022a} & Uncertainty rectified pyramid consistency \\
		&  {\fontfamily{cmtt}\selectfont SSLCCT} & Ouali et al., 2020~\cite{Ouali2020} & Cross-consistency training \\	
		&  {\fontfamily{cmtt}\selectfont SSLCPS} & Chen et al., 2021~\cite{Chen2021cps} & Cross-pseudo supervision  \\		\hline
		WSL  &  {\fontfamily{cmtt}\selectfont WSLEntropyMinimization} & - & Adapted from~\cite{Grandvalet2005} \\
		&  {\fontfamily{cmtt}\selectfont WSLTotalVariation} & - & Total variation loss for regularization \\
		&  {\fontfamily{cmtt}\selectfont WSLMumfordShah} & Kim et al. 2020~\cite{Kim2020} & Mumford-shah loss for regularization \\
		&  {\fontfamily{cmtt}\selectfont WSLGatedCRF} & Obukhov et al., 2019~\cite{Obukhov2019} & Gated CRF for regularization \\
		&  {\fontfamily{cmtt}\selectfont WSLUSTM} & Liu et al., 2022~\cite{Liu2022pr} & Adapt UAMT with transform-consistency \\
		&  {\fontfamily{cmtt}\selectfont WSLDMPLS} & Luo et al., 2022~\cite{Luo2022miccai} & Dynamically mixed pseudo label supervision \\
		\hline
		NLL  &  {\fontfamily{cmtt}\selectfont NoiseRobustDiceLoss} & Wang et al., 2020~\cite{Wang2020c} & Train with {\fontfamily{cmtt}\selectfont SegmentationAgent} \\ 
		&  {\fontfamily{cmtt}\selectfont GCELoss} & Zhang et al., 2018~\cite{Zhang2018d} & Train with {\fontfamily{cmtt}\selectfont SegmentationAgent} \\ 
		&  {\fontfamily{cmtt}\selectfont MAELoss} & Ghosh et al., 2017~\cite{Ghosh2017} & Train with {\fontfamily{cmtt}\selectfont SegmentationAgent} \\ 
		&  {\fontfamily{cmtt}\selectfont NLLCoTeaching} & Han et al., 2018~\cite{Han2018} & Co-teaching between two networks \\ 
		&  {\fontfamily{cmtt}\selectfont NLLCLSLSR} & Zhang et al., 2020~\cite{Zhang2020i} & Confident learning with spatial label smoothing \\ 
		&  {\fontfamily{cmtt}\selectfont NLLTriNet} & Zhang et al., 2020~\cite{Zhang2020trinet} & Tri-network combined with sample selection  \\ 
		&  {\fontfamily{cmtt}\selectfont NLLDAST} & Yang et al., 2022~\cite{Yang2022jbhi} & Divergence-aware selective training \\ 
		\hline
	\end{tabular}
	\label{tab:anno-eff-methods}.
\end{table*}

With the above advanced functionalities, three derivatives of  {\fontfamily{cmtt}\selectfont SegmentationAgent} have been currently implemented in PyMIC for semi-supervised learning, weakly supervised learning and noisy label learning, respectively: {\fontfamily{cmtt}\selectfont SSLSegAgent}, {\fontfamily{cmtt}\selectfont WSLSegAgent} and {\fontfamily{cmtt}\selectfont NLLSegAgent}, as shown in Fig.~\ref{fig:overview}. Then,  different kinds of specific annotation-efficient learning methods are derived from one of these  abstract classes. 
Table~\ref{tab:anno-eff-methods} shows the detailed annotation-efficient learning methods currently implemented in PyMIC.

The  {\fontfamily{cmtt}\selectfont SSLSegAgent} is an abstract agent for SSL and provides the general setting of a combination of annotated and unannotated subsets for training. It is the parent class of different semi-supervised segmentation methods, such as {\fontfamily{cmtt}\selectfont SSLMeanTeacher}~\cite{Tarvainen2017} based on mean teacher and {\fontfamily{cmtt}\selectfont SSLURPC}~\cite{Luo2022a} that uses uncertainty rectified pyramid consistency. Other methods including entropy minimization~\cite{Grandvalet2005}, UAMT~\cite{Yu2019}, Cross-Pseudo Supervision (CPS)~\cite{Chen2021cps} and Cross-Consistency Training (CCT)~\cite{Ouali2020} are also implemented as child classes of  {\fontfamily{cmtt}\selectfont SSLSegAgent}.

For weakly supervised segmentation, PyMIC supports learning from partial annotations (e.g., scribbles), and a {\fontfamily{cmtt}\selectfont PartialCELoss} is implemented for computing cross entropy loss only for the annotated pixels. It is combined with different regularization losses for training in derivatives of {\fontfamily{cmtt}\selectfont WSLSegAgent}, such as {\fontfamily{cmtt}\selectfont WSLEntropyMinimization}~\cite{Grandvalet2005}, {\fontfamily{cmtt}\selectfont WSLTotalVariation}, {\fontfamily{cmtt}\selectfont WSLMumfordShah}~\cite{Kim2020} and {\fontfamily{cmtt}\selectfont WSLGatedCRF}~\cite{Obukhov2019}. These agents can reuse networks for supervised learning during training.  In addition, some WSL methods~\cite{Liu2022pr,Luo2022miccai} require modified neural networks or multiple forward passes in each iteration, and PyMIC implemented their training procedures accordingly, such as  {\fontfamily{cmtt}\selectfont WSLUSTM}~\cite{Liu2022pr} that combines UAMT~\cite{Yu2019} with transform consistency for regularizing predictions of unannotated pixels, and
{\fontfamily{cmtt}\selectfont WSLDMPLS}~\cite{Luo2022miccai} that dynamically mixes predictions at two decoders as the pseudo label for training with scribble annotations. 

For noisy label learning, PyMIC has implementations of several noise-robust loss functions, such as {\fontfamily{cmtt}\selectfont GCELoss}~\cite{Zhang2018d}, {\fontfamily{cmtt}\selectfont MAELoss}~\cite{Ghosh2017} and  {\fontfamily{cmtt}\selectfont NoiseRobustDiceLoss}~\cite{Wang2020c}. They can be used by {\fontfamily{cmtt}\selectfont SegmentationAgent} for training with a standard supervised learning paradigm. In addition, some NLL methods require special training procedures like interaction between multiple networks, sample selection and cross-supervision. Child classes of {\fontfamily{cmtt}\selectfont NLLSegAgent} are implemented to support them, such as {\fontfamily{cmtt}\selectfont NLLCoTeaching}~\cite{Han2018}, {\fontfamily{cmtt}\selectfont NLLCLSLSR}~\cite{Zhang2020i} that combines confident learning with spatial label smoothing, {\fontfamily{cmtt}\selectfont NLLTriNet}~\cite{Zhang2020trinet} that uses three networks for training, and {\fontfamily{cmtt}\selectfont NLLDAST}~\cite{Yang2022jbhi} that combines a dual-branch network with divergence-aware selective training.

\subsection{Flexibility and extendability}
The modular design of PyMIC provides high flexibility and extensibility to different levels of users. PyMIC supports configuration files for specifying the dataset, data augmentation methods, network, loss functions and training parameters such as batch size, optimizer and  learning rate scheduler. Users are allowed to just edit the configuration file without writing code to run the training and inference pipeline with different settings. In addition, different modules in PyMIC can be customized by a user, so that the user only needs to define and set the customized module in the segmentation agent without re-implementing the remaining parts. For example, a user may define a customized network by inheriting {\fontfamily{cmtt}\selectfont nn.Module} and use {\fontfamily{cmtt}\selectfont SegmentationAgent.set\_network()} to easily integrate it into the existing pipeline.

\subsection{Platform and dependencies}
PyMIC is implemented in Python (Python Software Foundation, Wilmington, DA, U.S., version 3.6 or higher). It depends on PyTorch, Numpy, Pandas, SciPy and SimpleITK. The detailed dependent libraries and versions are encoded in a {\fontfamily{cmtt}\selectfont requirements.txt} file in the code repository, which enables automatic installation of compatible libraries. Users can easily install PyMIC via the {\fontfamily{cmtt}\selectfont pip} installation tool, i.e., running the command of {\fontfamily{cmtt}\selectfont pip install PYMIC}. The documentation  is hosted on Read the Docs\footnote{https://pymic.readthedocs.io/en/latest/},  which facilitates users starting with PyMIC quickly.

\begin{figure*}
	\centering
	\includegraphics[width=1.0\linewidth]{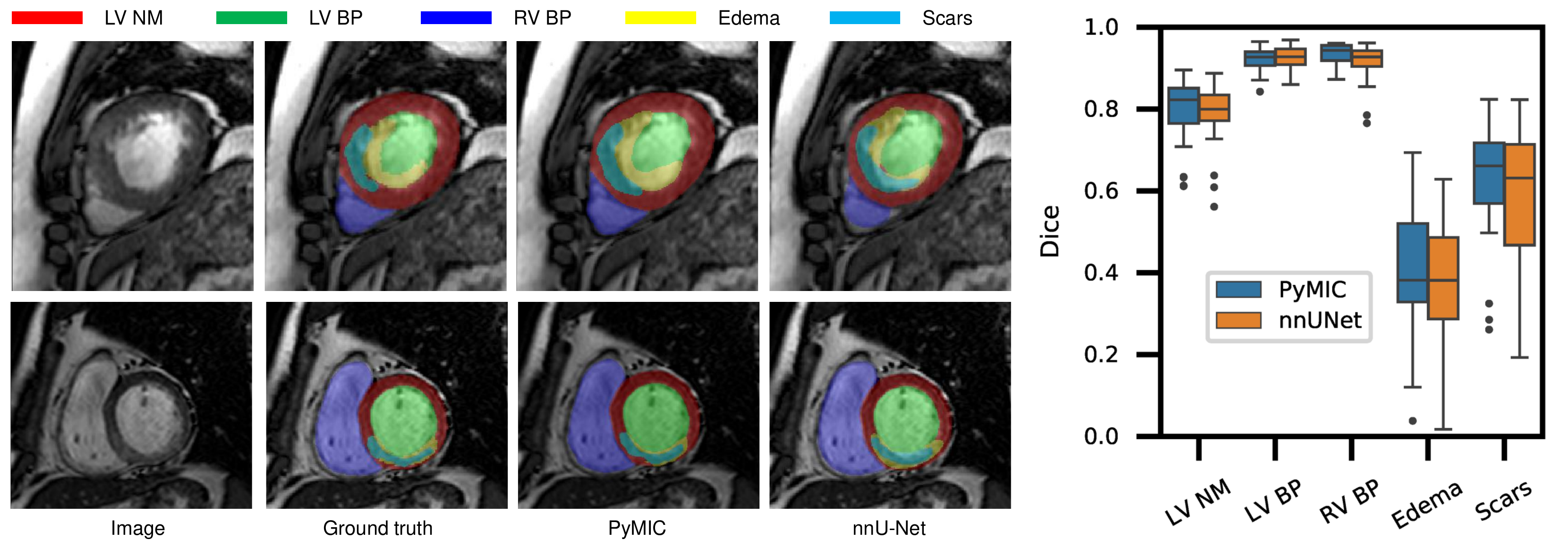}
	\caption{Myocardial pathology segmentation results obtained by PyMIC and nnU-Net~\cite{Isensee2021}. Left: qualitative comparison. Right: quantitative evaluation in terms of Dice. LV: Left ventricle; RV: Right ventricle; NM: Normal myocardium; BP: Blood pool.}
	\label{fig:myops}
\end{figure*}
\section{Results: Applications of PyMIC}
In this section, we present several user cases of PyMIC for 2D and 3D medical image segmentation with full annotations, partial annotations, weak annotations and noisy annotations, respectively. These user cases are available on GitHub\footnote{https://github.com/HiLab-git/PyMIC\_examples}. The experiments were conducted on a Ubuntu desktop with two NVIDIA GTX 1080 Ti GPUs available.

\subsection{Fully supervised segmentation}
Based on PyMIC, we have previously won two medical image segmentation challenges hold on MICCAI conference: MyoPS 2020~\cite{Lilei2022} for myocardial pathology segmentation and CATARACTS 2020~\cite{Luengo2021} for automatic tool annotation for cataRACT surgery. Here we demonstrate how PyMIC can be easily used for these two fully supervised segmentation tasks.

The MyoPS 2020 challenge aims to segment five structures from Cardiac Magnetic Resonance (CMR) images: Left Ventricular (LV) Normal Myocardium (NM), LV Blood Pool (BP), Right Ventricular (RV) BP, LV myocardial edema and scars. It provides 45 cases of three-sequence CMR images (bSSFP, LGE and T2 CMR), where the sequences in each case had been aligned and resampled to the same resolution by the organizers. Each case had 2-6 slices with inter-slice spacing 12.0-23.0~mm. The matrix size was around 480 $\times$ 480 with pixel size around 0.73~mm $\times$ 0.73~mm. Due to the small slice number, we used 2D networks for experiments with this dataset. The images was split into 25 cases for training and 20 cases for testing. As the labels of the testing cases are not publicly available, we use the official training cases for experiments, where 5-fold cross validation on the 25 cases is conducted in this paper.

We used {\fontfamily{cmtt}\selectfont SegmentationAgent} with a configuration file for training and inference, where a  {\fontfamily{cmtt}\selectfont UNet2D} was trained by a combination of  {\fontfamily{cmtt}\selectfont DiceLoss} and  {\fontfamily{cmtt}\selectfont CrossEntropyLoss}. A lot of {\fontfamily{cmtt}\selectfont Transform} objects, including {\fontfamily{cmtt}\selectfont RandomRotate, RandomCrop, RandomFlip, NormalizeWithMeanStd, GammaCorrection, GaussianNoise}, were used for data augmentation. The {\fontfamily{cmtt}\selectfont Adam} optimizer was used with {\fontfamily{cmtt}\selectfont ReduceLROnPlateau} scheduler for training, with initial learning rate $0.001$ and maximal iteration number 30$k$. During inference, test time augmentation was used in {\fontfamily{cmtt}\selectfont Inferer}. The quantitative evaluation metrics were computed using PyMIC's {\fontfamily{cmtt}\selectfont evaluation} method and aggregated over all folds. 

The segmentation results by PyMIC was compared with those by nnU-Net\footnote{https://github.com/MIC-DKFZ/nnUNet}~\cite{Isensee2021}.  As shown in 
Fig.~\ref{fig:myops}, PyMIC performed better than nnU-Net for the segmentation task. The average Dice value obtained by PyMIC was 0.795, 0.922, 0.935, 0.400, 0.613 for LV NM, LV BP, RV BP, edema and scars, respectively. Note that the edema and scars are small objects that are hard to segment, some advanced strategies such as model ensemble and coarse-to-fine segmentation can be combined with PyMIC for better performance, as demonstrated in~\cite{Zhai2020}.

The CATARACTS 2020~\cite{Luengo2021} challenge is about semantic  segmentation of cataract surgery videos. There were three sub-tasks for segmentation of different numbers of classes, and in this paper we deal with Task I where 8 classes need to be segmented: 4 for anatomy (pupil, iris, cornea and skin), 1 for all instruments and 3 for other objects (surgical tap, eye retractors and hand). The challenge dataset contains 4670 images from 25 videos for training and 531 images from 10 videos for testing. As the labels of testing images are not publicly available, we only used the official training set for experiment, and it was split into 3550 (19 videos), 534 (3 videos) and 586 (3 videos) for training, validation and testing, respectively.

\begin{figure*}
	\centering
	\includegraphics[width=0.85\linewidth]{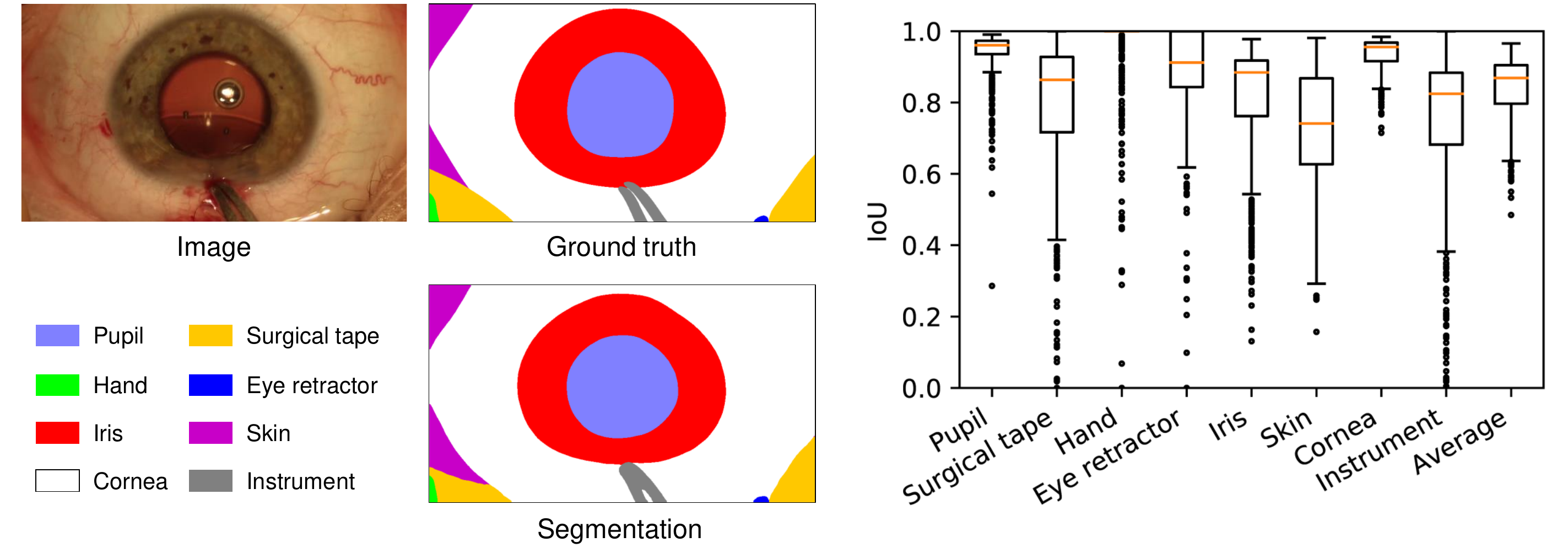}
	\caption{Semantic segmentation  of cataract surgery images  obtained by PyMIC. Left: qualitative comparison. Right: quantitative evaluation in terms of IoU. }
	\label{fig:cataract}
\end{figure*}

\begin{table*}[h]
	\caption{Quantitative comparison of different semi-supervised methods for cardiac MRI segmentation on the ACDC dataset, with 10\% of training images annotated. $^*$ denotes significant improvement from the baseline ($p$-value < 0.05) based on a paired t-test.}
	\centering
	\small 
	\label{tab:acdc_ssl}
	\begin{tabular}{ l | llll | llll }
		\hline
		\multirow{2}{*}{Methods} & \multicolumn{4}{c|}{Dice (\%)} & \multicolumn{4}{c }{HD95 (mm)} \\
		\cline{2-5} \cline{6-9} 
		&  RV & Myo & LV & Average & RV & Myo & LV & Average \\ \hline
		Baseline & 83.32$\pm$11.91&  85.93$\pm$5.30 & 91.76$\pm$5.87 & 87.00$\pm$6.39& 11.34$\pm$10.72 & 7.64$\pm$8.15 & 7.27$\pm$8.66 & 8.75$\pm$7.35 \\ 
		EM~\cite{Grandvalet2005} & 84.25$\pm$11.47& 86.35$\pm$5.45 & \textbf{92.96$\pm$4.53}$^*$ & 87.86$\pm$5.97 & 12.89$\pm$17.69 & 5.75$\pm$6.69$^*$ & 6.62$\pm$8.23 & 8.42$\pm$7.87 \\ 
		UAMT~\cite{Yu2019} & 86.90$\pm$8.73$^*$ & 86.23$\pm$4.54 & 92.38$\pm$5.01 & 88.50$\pm$4.62$^*$ & 10.60$\pm$12.15 & 6.12$\pm$6.77 & 7.03$\pm$6.85 & 7.92$\pm$6.27 \\ 
		URPC~\cite{Luo2022a} &  88.70$\pm$7.20$^*$ & 87.09$\pm$4.73$^*$ &  92.78$\pm$4.78 & 89.53$\pm$4.31$^*$ & 8.81$\pm$10.96$^*$ & 5.73$\pm$6.84$^*$ & \textbf{5.63$\pm$6.62} & 6.73$\pm$6.24$^*$ \\ 
		CCT~\cite{Ouali2020} & 85.98$\pm$10.75$^*$ & 86.59$\pm$5.00$^*$ & 91.98$\pm$5.42 & 88.18$\pm$5.48$^*$ & 9.24$\pm$8.03 & 6.28$\pm$7.75 & 7.07$\pm$8.32 & 7.53$\pm$5.92 \\ 
		CPS~\cite{Chen2021cps} & \textbf{89.09$\pm$7.32}$^*$ & \textbf{87.09$\pm$4.44}$^*$ & 92.49$\pm$5.18 & \textbf{89.56$\pm$4.11}$^*$ & \textbf{6.78$\pm$6.56}$^*$ & \textbf{5.02$\pm$4.96}$^*$ & 6.42$\pm$8.01 & \textbf{6.08$\pm$4.87}$^*$ \\ 
		
		\hline
	\end{tabular}
\end{table*}
For this task we combined PyMIC with a customized CNN, where the network in {\fontfamily{cmtt}\selectfont SegmentationAgent}  was set as a pre-trained {\fontfamily{cmtt}\selectfont HRNet}~\cite{Wang2019g} implemented in a third-party repository\footnote{https://github.com/HRNet/HRNet-Semantic-Segmentation}. The data augmentation methods were similar to those used in the MyoPS challenge, and we replaced {\fontfamily{cmtt}\selectfont RandomRotate} by  {\fontfamily{cmtt}\selectfont RandomRescale}. We used a batch size of 12 and trained {\fontfamily{cmtt}\selectfont HRNet} on two GPUs with {\fontfamily{cmtt}\selectfont DiceLoss} and  {\fontfamily{cmtt}\selectfont CrossEntropyLoss}. The initial learning rate was $10^{-4}$ with maximal iteration number of $100k$. Test time augmentation and sliding window inference were enabled in {\fontfamily{cmtt}\selectfont Inferer}. The segmentation results are shown in Fig.~\ref{fig:cataract}, where the left side shows a qualitative comparison between the segmented objects and the ground truth, and the right side shows quantitative evaluation in terms of Intersection over Union (IoU) as used in~\cite{Luengo2021}.

\subsection{Semi-supervised segmentation}\label{sec:ssl_exp}
To demonstrate semi-supervised segmentation in PyMIC, we first used the Automatic Cardiac Diagnosis Challenge (ACDC) dataset~\cite{Bernard2018}  containing 200 short-axis cardiac cine-MR volumes of 100 patients for segmentation of the Right Ventricle (RV), Myocardium (Myo) and Left Ventricle (LV). Each volume has 6-18 slices, with an inter-slice spacing of 5-10~mm. We randomly selected 20 volumes of 10 patients for validation, and 40 volumes of 20 patients for testing. The remaining 140 volumes of 70 patients were used for training, where 14 volumes from 7 patients (10\%) and the other 126 volumes were used as the annotated and unannotated samples, respectively. 

We compared several SSL methods implemented in PyMIC with the baseline of learning only from the annotated samples: {\fontfamily{cmtt}\selectfont SSLEntropyMinimization}~\cite{Grandvalet2005}, {\fontfamily{cmtt}\selectfont SSLUAMT}~\cite{Yu2019}, {\fontfamily{cmtt}\selectfont SSLURPC}~\cite{Luo2022a}, {\fontfamily{cmtt}\selectfont SSLCCT}~\cite{Ouali2020}, and {\fontfamily{cmtt}\selectfont SSLCPS}~\cite{Chen2021cps}.
All these methods used a 2D network ({\fontfamily{cmtt}\selectfont UNet2D}) as the backbone due to the small number of slices and large inter-slice spacing in the volumetric images, and were trained with a batch size of 8 that consists of 4 annotated and 4 unannotated samples. Adam optimizer was used for training, with initial learning rate of $10^{-3}$ and maximal iteration number of $30k$. The results were post-processed by selecting the largest connected component implemented in {\fontfamily{cmtt}\selectfont post\_process} module of PyMIC.  Quantitative evaluation results in terms of Dice and HD95 are shown in Table~\ref{tab:acdc_ssl}. The baseline method obtained an average Dice of 83.32\%, 85.93\% and 91.76\% for RV, Myo and LV, respectively. The SSL methods significantly improved the performance, e.g., CPS~\cite{Chen2021cps} improved the average Dice to 89.09\%, 87.09\% and 92.49\% for the three classes, respectively.

\begin{table}
	\caption{Quantitative comparison of different semi-supervised methods for 3D left atrial segmentation, with 10\% of training images annotated. $^*$ denotes significant improvement from the baseline ($p$-value < 0.05) based on a paired t-test.}
	\centering
	\small 
	\label{tab:semi-atrial}
	\begin{tabular}{ l | lll}
		\hline
		
		&  Dice (\%) & HD95 (mm) & ASSD (mm) \\ \hline
		Baseline &  82.58$\pm$8.62 &   20.19$\pm$14.75 & 3.62$\pm$2.96  \\ 
		EM~\cite{Grandvalet2005} & 84.62$\pm$9.14 &  15.75$\pm$13.74 & 2.83$\pm$2.87$^*$ \\ 
		UAMT~\cite{Yu2019} &  84.85$\pm$6.97 &  24.00$\pm$18.73 & 3.93$\pm$3.17\\ 
		URPC~\cite{Luo2022a} &  85.06$\pm$8.01$^*$ &  18.44$\pm$13.99 & 3.23$\pm$2.83 \\
		CPS~\cite{Chen2021cps} &  \textbf{86.50$\pm$4.68}$^*$ &  \textbf{13.36$\pm$12.92}$^*$ &  \textbf{2.39$\pm$2.30}$^*$ \\
		\hline
	\end{tabular}
\end{table}
\begin{figure*}
	\centering
	\includegraphics[width=1.0\linewidth]{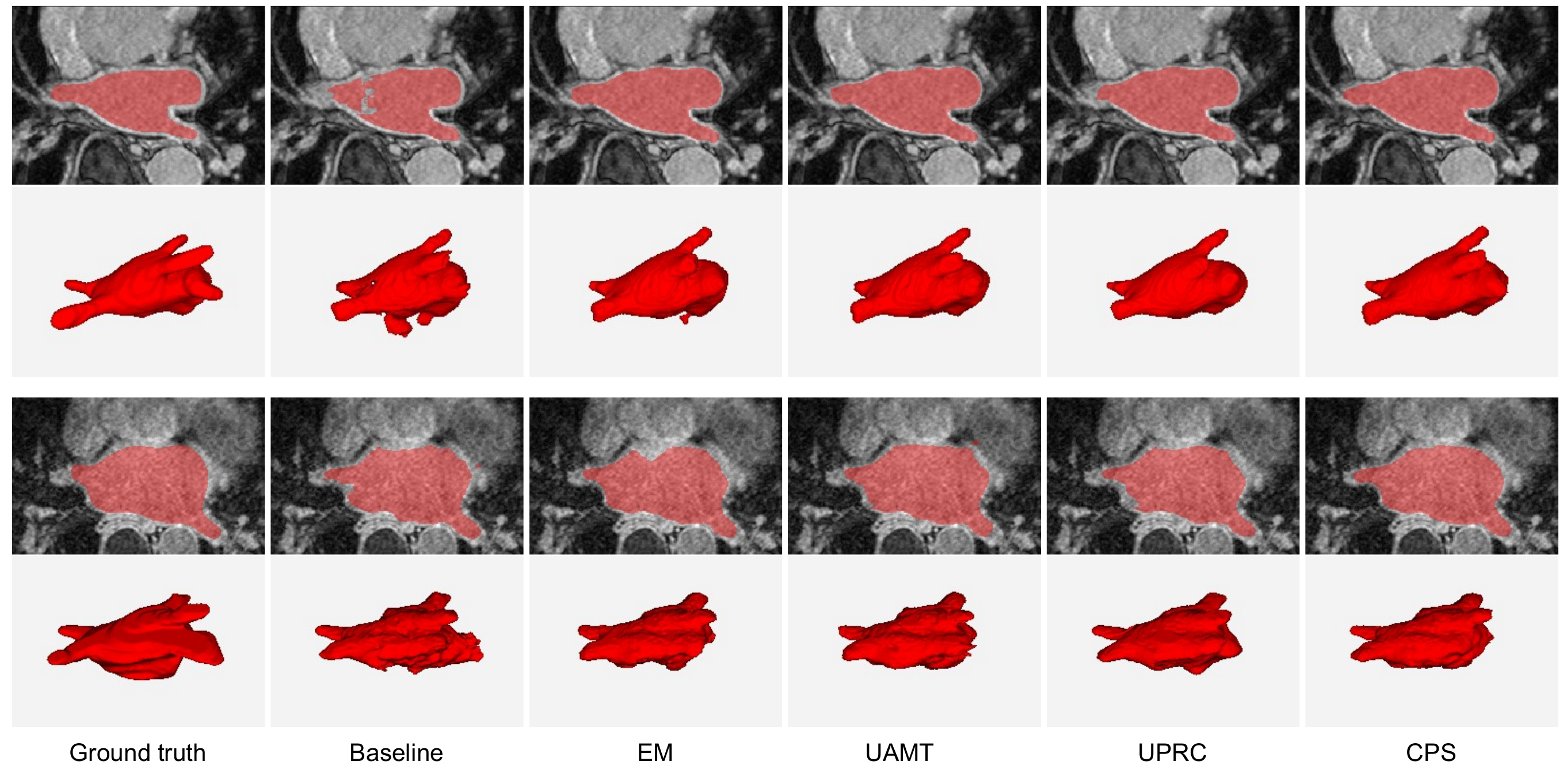}
	\caption{Semi-supervised 3D left atrial segmentation results. The models were trained with 7 annotated and 65 unannotated volumes. The first two rows show the segmentation results of one patient with 2D and 3D visualizations, respectively, and the last two rows show the results of another patient.  }
	\label{fig:atrial_ssl}
\end{figure*}

In addition, we also demonstrate PyMIC's support for 3D semi-supervised segmentation with the left atrial dataset~\cite{Xiong2020} that consists of 100 3D atrial gadolinium-enhanced MR images. The original images have a resolution of 0.625 $\times$ 0.625 $\times$ 0.625~mm$^3$, with spatial dimensions of 576 $\times$ 576 $\times$ 88 or 640 $\times$ 640 $\times$ 88 pixels. For preprocessing, we cropped the images into a dimension of 256 $\times$ 160 $\times$ 88. The dataset was split into 72, 8 and 20 cases for training, validation and testing, respectively. For the 72 training images, we took 7 ($\sim$10\%) as annotated images and the other 65 ($\sim$90\%) as unannotated images. {\fontfamily{cmtt}\selectfont UNet3D}~\cite{Abdulkadir2016} was used as the backbone network, and we compared the baseline of learning only from the annotated images with: {\fontfamily{cmtt}\selectfont SSLEntropyMinimization}~\cite{Grandvalet2005}, {\fontfamily{cmtt}\selectfont SSLUAMT}~\cite{Yu2019}, {\fontfamily{cmtt}\selectfont SSLURPC}~\cite{Luo2022a},  and {\fontfamily{cmtt}\selectfont SSLCPS}~\cite{Chen2021cps}.  Note that {\fontfamily{cmtt}\selectfont SSLCCT}~\cite{Ouali2020} was not used in this experiment due to the limit of memory. The patch size was 112 $\times$ 96 $\times$ 72, with a batch size of 4  consisting of 2 annotated and 2 unannotated samples. We also used Adam optimizer for training, with initial learning rate of $10^{-3}$ and maximal iteration number of $20k$.

Table~\ref{tab:semi-atrial} shows the quantitative evaluation results of these compared methods. The baseline obtained an average Dice of 82.58\%, and CPS~\cite{Chen2021cps} improved it to 86.50\%, which outperformed the other semi-supervised learning methods. CPS also obtained the lowest HD95 and ASSD  among the compared methods. URPC~\cite{Luo2022a} ranked the second place in terms of average Dice, and  EM~\cite{Grandvalet2005} ranked the second place in terms of HD95 and ASSD. Fig.~\ref{fig:atrial_ssl} shows a visual comparison between these different methods. It can be observed that the baseline  obtained obvious under-segmentation and over-segmentation, while the semi-supervised methods improved the segmentation quality by leveraging unannotated images for learning.

\subsection{Learning from sparse annotations}
\begin{figure*}
	\centering
	\includegraphics[width=1.0\linewidth]{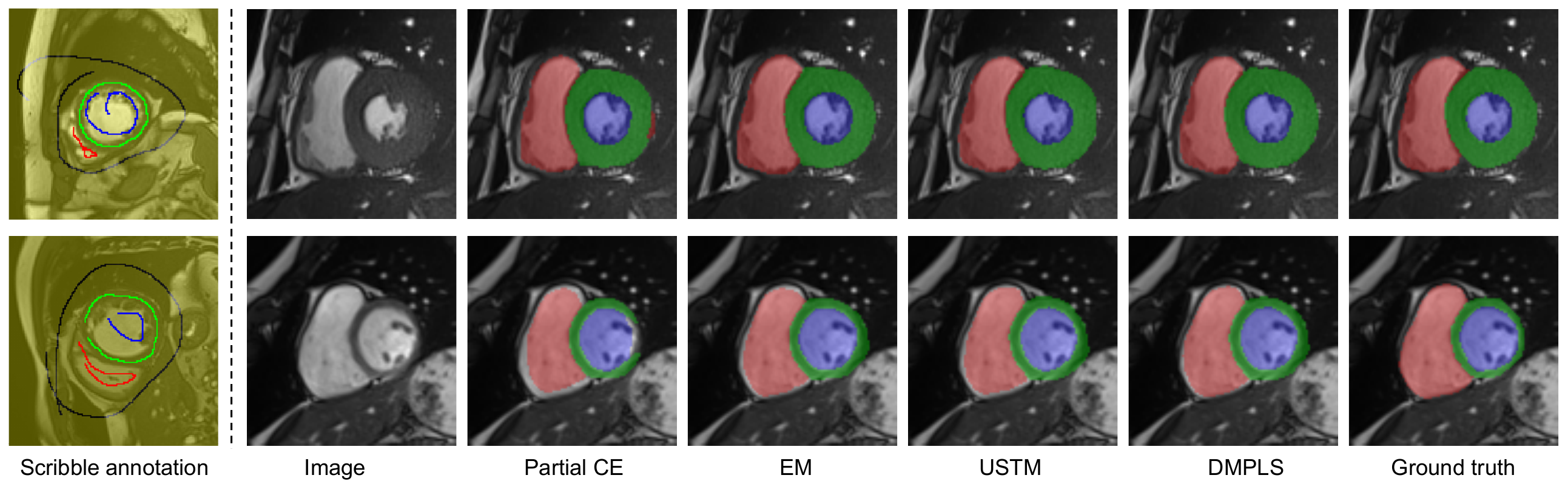}
	\caption{Weakly supervised cardiac structure segmentation on ACDC dataset.}
	\label{fig:acdc_wsl}
\end{figure*}

We also used the ACDC dataset as an example for weakly supervised segmentation, where scribble annotations provided by Valvano et al.~\cite{Valvano2021} instead of full annotations were used for training, as shown in the left side of Fig.~\ref{fig:acdc_wsl}. The data split, backbone network, optimizer, learning rate and post-process method were the same as in Section~\ref{sec:ssl_exp}. Five WSL methods implemented in PyMIC were evaluated: {\fontfamily{cmtt}\selectfont WSLEntropyMinimization}, {\fontfamily{cmtt}\selectfont WSLTotalVariation}, {\fontfamily{cmtt}\selectfont WSLGatedCRF~\cite{Obukhov2019}}, {\fontfamily{cmtt}\selectfont WSLUSTM~\cite{Liu2022pr}} and {\fontfamily{cmtt}\selectfont WSLDMPLS}~\cite{Luo2022miccai}, and they were compared with the baseline of learning only from the annotated pixels with {\fontfamily{cmtt}\selectfont PartialCELoss}. Quantitative evaluation of these methods are shown in Table~\ref{tab:acdc_wsl}. Using partial CE loss only obtained an average Dice of 86.39\%, and DMPLS~\cite{Luo2022miccai} improved it to 89.28\%, which outperformed the other compared methods. Note that the performance of DMPLS  implemented in this paper is better than that in the original paper~\cite{Luo2022miccai}, showing the effectiveness of PyMIC. The right side of Fig.~\ref{fig:acdc_wsl} shows qualitative comparisons of these methods. It can observed that the partial CE baseline obtained noisy segmentation in the first row and under-segmentation in the second row respectively, while the other weakly supervised methods obtained better results, and DMPLS outperformed the other compared methods.

\begin{table*}
	\caption{Quantitative comparison of different methods learning from scribbles for cardiac MRI segmentation on the ACDC dataset. $^*$ denotes significant improvement from the baseline ($p$-value < 0.05) based on a paired t-test.}
	\centering
	\small 
	\label{tab:acdc_wsl}
	\begin{tabular}{ l | llll | llll }
		\hline
		\multirow{2}{*}{Methods} & \multicolumn{4}{c|}{Dice (\%)} & \multicolumn{4}{c }{HD95 (mm)} \\
		\cline{2-5} \cline{6-9} 
		&  RV & Myo & LV & Average & RV & Myo & LV & Average \\ \hline
		Partial CE &  85.96$\pm$5.37 &  83.66$\pm$5.87 & 89.54$\pm$7.79 & 86.39$\pm$4.86 & 7.30$\pm$4.25  & 7.08$\pm$8.16 & 6.84$\pm$7.55 & 7.07$\pm$5.15 \\ 
		EM~\cite{Grandvalet2005} &  87.83$\pm$7.19$^*$ & 85.93$\pm$4.31$^*$ & 92.42$\pm$5.01$^*$ & 88.73$\pm$4.00$^*$ & 6.10$\pm$4.01$^*$ & \textbf{4.07$\pm$4.29}$^*$ & 4.29$\pm$4.66$^*$ & 4.82$\pm$3.39$^*$ \\ 
		TV &  87.93$\pm$6.17$^*$ & 85.84$\pm$4.11$^*$ & \textbf{92.71$\pm$4.13}$^*$ & 88.83$\pm$3.27$^*$ & 6.68$\pm$6.11 & 4.26$\pm$4.05$^*$ & 5.07$\pm$10.56 & 5.34$\pm$5.01 \\ 
		Gated CRF~\cite{Obukhov2019} &  86.94$\pm$5.77$^*$ & 85.40$\pm$4.47$^*$ & 91.79$\pm$5.44$^*$ & 88.05$\pm$3.86$^*$ & 6.49$\pm$3.69 & 4.73$\pm$4.65$^*$ & 5.35$\pm$6.65 & 5.53$\pm$3.78$^*$ \\ 
		USTM~\cite{Liu2022pr} &   87.00$\pm$6.59$^*$ & 84.74$\pm$4.83$^*$ &  91.84$\pm$4.42$^*$ & 87.86$\pm$3.80$^*$ & 6.87$\pm$4.73 & 4.74$\pm$3.99$^*$ & 4.27$\pm$3.29$^*$ & 5.30$\pm$2.78$^*$\\ 
		DMPLS~\cite{Luo2022miccai} &  \textbf{89.02$\pm$5.28}$^*$ & \textbf{86.17$\pm$4.50}$^*$ &  92.66$\pm$4.22$^*$ & \textbf{89.28$\pm$3.08}$^*$ &  \textbf{5.21$\pm$3.46}$^*$  & 4.26$\pm$3.89$^*$ & \textbf{3.97$\pm$4.09}$^*$ & \textbf{4.48$\pm$2.96}$^*$ \\
		\hline
	\end{tabular}
\end{table*}

\begin{figure}
	\centering
	\includegraphics[width=1.0\linewidth]{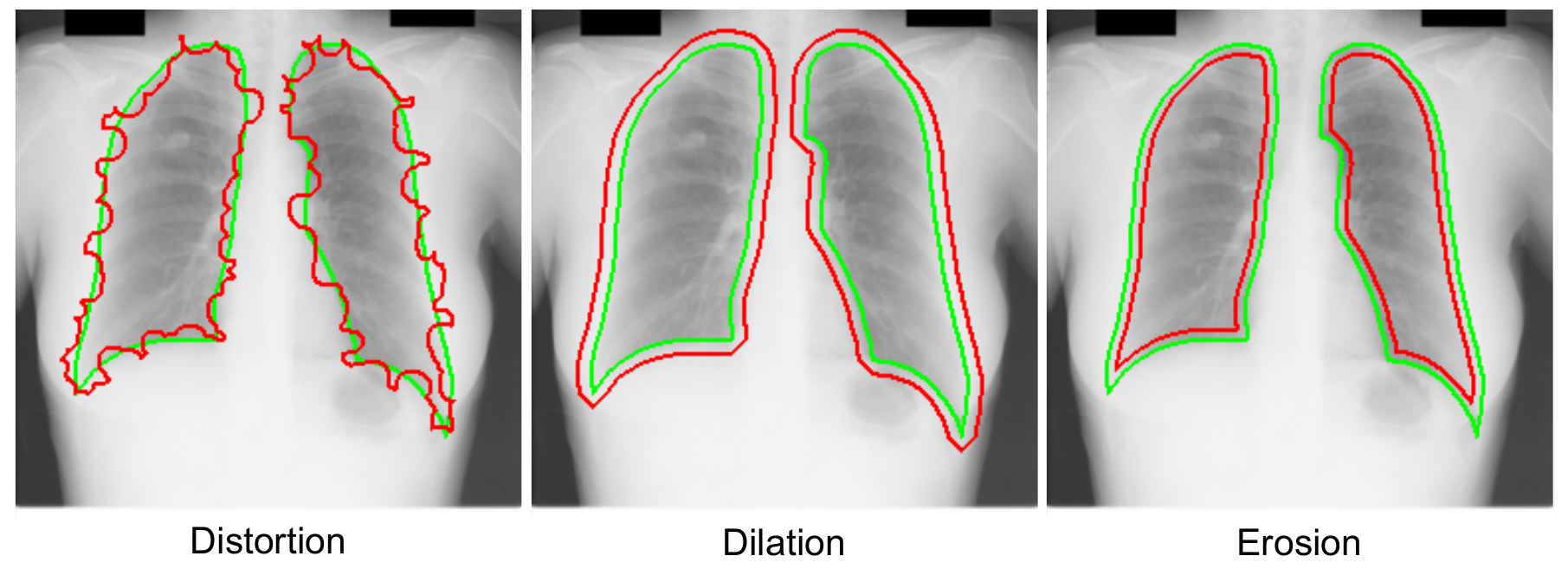}
	\caption{Examples of simulated noisy labels (red) compared with clean label (green) for lung segmentation.}
	\label{fig:jsrt_noise}
\end{figure}
\subsection{Dealing with noisy labels}
The JSRT dataset~\cite{Shiraishi2000} with 247 chest radiographs was used for experiments with NLL. The images have a size of 2048$\times$2048, and were resampled to 256$\times$256. We split the dataset into 180, 20 and 47 for training, validation and testing, respectively. We only consider the lung for segmentation, and simulated noisy annotations for 95\% of the training images (171 cases) by applying dilation, erosion or edge distortion to the ground truth clean labels, as shown in Fig.~\ref{fig:jsrt_noise}. 

\begin{table}
	\caption{Quantitative comparison of different methods learning from noisy labels for lung segmentation. $^\dagger$ denotes the best value that is significantly better than all the others ($p$-value < 0.05) based on a paired t-test.} 
	\centering
	\small 
	\label{tab:lung_nll}
	\begin{tabular}{ l | lll}
		\hline
		
		&  Dice (\%) & HD95 (pix) & ASSD (pix) \\ \hline
		Baseline (CE) &  86.87$\pm$3.43 &   25.57$\pm$30.00 & 6.83$\pm$3.06  \\ 
		GCE loss~\cite{Zhang2018d} & 87.87$\pm$3.31 &  14.05$\pm$10.94 & 5.58$\pm$1.68 \\ 
		CL-SLSR~\cite{Zhang2020i} &  89.52$\pm$3.22 &  12.19$\pm$11.29 & 4.50$\pm$1.35\\ 
		Co-teaching~\cite{Han2018} &  94.25$\pm$2.45 &  11.92$\pm$8.08 & 2.60$\pm$1.13 \\
		TriNet~\cite{Zhang2020trinet} &  94.87$\pm$1.62 &  \textbf{7.82$\pm$3.87}$^\dagger$ & 2.26$\pm$0.66 \\
		DAST~\cite{Yang2022jbhi} &  \textbf{96.94$\pm$1.57}$^\dagger$ &  11.85$\pm$13.74 & \textbf{1.84$\pm$1.51} \\
		\hline
	\end{tabular}
\end{table}
We compared six methods implemented in PyMIC for this NLL task: 1) the baseline of training with cross entropy loss and standard setting in 
{\fontfamily{cmtt}\selectfont SegmentationAgent}; 2) training with {\fontfamily{cmtt}\selectfont GCELoss}~\cite{Zhang2018d}; 3) Confident Learning with Spatial Label Smoothing Regularization (CL-SLSR)~\cite{Zhang2020i} implemented in {\fontfamily{cmtt}\selectfont NLLCLSLSR}; 4) {\fontfamily{cmtt}\selectfont NLLCoTeaching}~\cite{Han2018}, 5) {\fontfamily{cmtt}\selectfont NLLTriNet}~\cite{Zhang2020trinet}; and 6) NLLDAST~\cite{Yang2022jbhi}. All these methods used {\fontfamily{cmtt}\selectfont UNet2D}~\cite{Ronneberger2015} as the backbone network for segmentation, and were trained with a batch size of 8, Adam optimizer, initial learning rate of $10^{-3}$ and  iteration number of 10$k$. No post-processing was used at inference time. 
Quantitative evaluation results are listed in Table ~\ref{tab:lung_nll}. It shows that the baseline method obtained an average Dice of 86.87\%. GCE loss and CL-SLSR improved it to 87.87\% and 89.52\%, respectively. Co-teaching, TriNet and DAST obtained a higher performance improvement, with average Dice of 94.25\%, 94.87\% and 96.94\%, respectively, showing the effectiveness of dealing with noisy labels in PyMIC.

\section{Discussion and Conclusions}
The current success of deep learning for medical image segmentation mostly relies on a large annotated dataset for fully supervised learning, which is well supported by several existing deep learning toolkits for medical image computing~\cite{Isensee2021,monai2022,Gibson2018,Duan2020}.
However, in a wider range of scenarios, the annotations of medical images are expensive and time-consuming to acquire, leading to demand for annotation-efficient training methods for deep learning models. Recent years have seen a rapid increase of researches on annotation-efficient learning algorithms~\cite{Tajbakhsh2019,Wang2021d}, but there is a lack of toolkits with built-in and modular functionalities for  fast implementation of annotation-efficient learning methods. 

We present PyMIC, a Python toolkit for annotation-efficient deep learning for medical image segmentation. PyMIC shares some similarity with MONAI~\cite{monai2022} with key differences.  MONAI is a PyTorch-based framework for deep learning for healthcare. In terms of supervised segmentation, both PyMIC and MONAI have domain-specialized modules for image I/O, data transforms, network structures, loss functions, evaluation metrics, and training and inference, etc. However, MONAI aims to support a wider range of healthcare data such as images, videos, tubular data and EEG signals, and it mainly covers training pipelines for fully supervised learning. In contrast, PyMIC is specialized for annotation-efficient medical image segmentation, and it focuses on functionalities for training with imperfect annotations, such as partial, sparse and noisy annotations, although it also contains basic modules to support fast development of competitive fully supervised segmentation models.  

With the built-in modules in PyMIC, users can easily run state-of-the-art annotation-efficient learning algorithms to train their own models by just editing a configuration file. PyMIC is easy to install, and the source code can be accessed from the Git repository\footnote{https://github.com/HiLab-git/PyMIC}. PyMIC provides a high flexibility to configure a training pipeline with 2D and 3D networks in different segmentation tasks. For example, for segmentation tasks on the ACDC and left atrial datasets in Section~\ref{sec:ssl_exp}, the different SSL methods implemented in PyMIC are independent of the network structures, and the user only needs to set the backbone network as a 2D or a 3D network in the two tasks, without changing the code for training. The WSL and NLL methods in PyMIC also allow users to freely set different network structures in the configure file. In addition, PyMIC supports customized modules, where the user can reuse most part of the training pipeline in PyMIC and only needs to focus on implementing the customized component, which has a potential for boosting researches and applications of annotation-efficient medical image segmentation algorithms.  

The current released version of PyMIC is v0.3, and it has been successfully used for several research projects in medical image segmentation. For example, in fully supervised learning, some novel networks such as CA-Net~\cite{Gu2020a}, MG-Net~\cite{Wang2020g} and COPLE-Net~\cite{Wang2020c}  have been developed on the top of PyMIC, and they have been applied to segmentation of organs at risks for radiotherapy~\cite{Liu2021b}, fetal brain and placenta~\cite{Gu2020a}, and COVID-19 lesions~\cite{Wang2020c,Yang2022jbhi}. Segmentation models implemented in PyMIC have won some MICCAI segmentation challenges, including MyoPS 2020~\cite{Zhai2020} and CATARACTS 2020~\cite{Luengo2021}. In addition, it has been used in researches on annotation-efficient learning, such as semi-supervised segmentation of pulmonary fibrosis from CT volumes~\cite{Wang2022tmi}, 
scribble-based weakly supervised learning~\cite{Luo2022miccai}, and noise-robust learning~\cite{Wang2020c, Yang2022jbhi}. 

There are also some limitations in the current version of PyMIC that are expected to be addressed in the future. First, the current built-in networks mainly include some UNet-like structures, and we are working on implementing more state-of-the-art networks in PyMIC to facilitate the users. Second, there are some other types of annotation-efficient learning approaches not considered in PyMIC currently, such as self-supervised learning approaches~\cite{Chen2019c,Zhu2020b} and domain adaption methods~\cite{Guan2021,Jinaghao2022} for medical image segmentation. Thirdly, despite that our framework achieved comparable and even better performance than the self-configuring nnU-Net~\cite{Isensee2021} in fully supervised learning (Fig.~\ref{fig:myops}), it requires the user's expertise to design a suitable set of configuration parameters. In addition, the semi- and weakly supervised learning pipelines usually have a lower training efficiency than fully supervised approaches, and their interpretability  need to be improved.  In the future, PyMIC will be extended to support more medical image computing tasks, such as image classification and object detection. 
It will be actively updated to support more state-of-the-art deep learning models and annotation-efficient learning methods for medical image computing.  

In summary, we present the open-source toolkit PyMIC to fill the gap between existing deep learning toolkits and annotation-efficient medical image segmentation. It's modular functionalities support fast development of medical image segmentation models with limited annotations based on the built-in implementations of semi-supervised, weakly-supervised and noisy-label learning algorithms. It's flexibility allows customized modules with minimized implementation burden on users, which facilitates researches and applications of deep learning for medical image computing with low annotation cost. 

\section{Declaration of Competing Interest}
The authors declare no conflicts of interest.
\section{Acknowledgments}
This work was supported by the National Natural Science Foundation of China under Grant 61901084 and 62271115, National Key Research and Development Program of China (2020YFB1711500), the 1·3·5 project for disciplines of excellence, West China Hospital, Sichuan University (ZYYC21004).

\bibliographystyle{model2-names}
\bibliography{pymic_bib}

\end{document}